\documentclass[12pt]{article}

\usepackage{amssymb}
\usepackage{color}
\usepackage{epsfig,amssymb,amsfonts,amsmath,graphicx,dsfont,cite,xfrac}
\usepackage{authblk}
\usepackage{subcaption}
\definecolor{mygray}{gray}{0.5}

\usepackage{cite}
\usepackage[colorlinks=true,linkcolor=blue,citecolor=red]{hyperref}

\newcommand{\be}{\begin{equation}}
\newcommand{\ee}{\end{equation}}
\newcommand{\bea}{\begin{eqnarray}}
\newcommand{\eea}{\end{eqnarray}}

\title{Position-dependent mass systems: Classical and quantum pictures\thanks{Extended abstract of  ``Algebraic approach to position-dependent mass systems in both classical and quantum pictures'', a series of three lectures delivered by the author in the {\em VIII School on Geometry and Physics}, 24 June--8 June 2019, organized by the Department of Mathematical Physics of the University of Bia\l ystok, in Bia\l owie\.za, Poland (\url{http://wgmp.uwb.edu.pl/wgmp38/part_s.html}).}}

\author[${}$]{Oscar Rosas-Ortiz}

\affil[${}$]{\footnotesize Physics Department, Cinvestav, AP 14-740, 07000
M\'exico City, Mexico}

\date{}
\begin{document}

\maketitle

\begin{abstract}
The present work is an extended abstract from a series of lectures addressed to introduce elements of the theory of position-dependent mass systems in both, classical and quantum mechanics.
\end{abstract}


\section{Motivation}

The study of systems endowed with position-dependent mass (PDM) is a subject of great interest in many branches of physics. Among others, the examples include dynamical systems in curved spaces with either constant curvature \cite{Koz92,Car12} or non-constant curvature \cite{Rag10}, geometric optics \cite{Wol04}, semiconductor theory \cite{Bas81,Bas82}, motion of rockets \cite{Som94}, raindrop problem \cite{Kra81}, variable mass oscillators \cite{Flo03}, inversion potential for NH3 in density theory \cite{Aqu98}, evolution of binary systems \cite{Haf67}, effects of galactic mass loss \cite{Ric82}, neutrino mass oscillations \cite{Bet86}, and the problem of a rigid body against a liquid free surface \cite{Pes03}. 

Despite the large number of models used to describe the dynamics of PDM systems, the principles of the underlying theory are not fully understood. In the classical picture a position-dependent mass function $m(x)$ gives rise to `forces quadratic in the velocity' which lead to nonlinear differential equations of motion in the Newtonian approach \cite{Math74,Cru13}. In turn, the Hermiticity of the Hamiltonian of a quantum system is part of the problem to solve if the mass depends on the position \cite{Cru09}. Nevertheless, the difficulties appearing in both pictures can be faced by using the factorization method together with Lie algebraic tools \cite{Cru13,Cru09,Cru07,Kur08,Cru08,Cru08b,Cru13b,Cru14,Scr16,Cru18,Cru11}.

\section{Classical picture}

The one-dimensional dynamical law for a system with position-dependent mass $m(x)>0$ that is acted by a force $F$ depending on position~$x$, velocity $\dot x$, and time $t$, may be written as \cite{Cru13}:
\begin{equation}
F(x, \dot x; t) = \frac{dp}{dt}= m'(x) \dot x^2 + m(x) \ddot x,
\label{force}
\end{equation}
where $p=m(x)\dot x$ is the linear momentum. Hereafter $\dot f$ and $f'$ stand for time and position derivatives of~$f$ respectively. Let us rewrite the Newton equation~\eqref{force} in the standard form
\begin{equation}
m(x) \ddot x = F_{\rm net}(x, \dot x; t) \equiv F(x, \dot x; t) - m'(x) \dot x^2.
\label{force2}
\end{equation}
We immediately see that the term quadratic in the velocity corresponds to the thrust of the system, so that Eq.~\eqref{force2} indicates how this term alters the velocity. Indeed, as $\dot x^2 \geq 0$, the system is accelerated (decelerated) if the rate $m'(x)$ is negative (positive). Thus, a particle suf\/fering a~spatial variation of its mass is acted by a net force $F_{\rm net}$ that results from the combination of the external force~$F$ and the thrust $-m'(x)\dot x^2$.

Applying the D'Alembert principle and assuming that the external force is derivable from a scalar potential function $\mathcal{V}(x)$ that does not depend on either velocity or time, from \eqref{force} we arrive at the Lagrange equation
\begin{equation}
\frac{d}{dt} \left( \frac{\partial L}{\partial \dot x} \right) - \frac{\partial L}{\partial x} =
\widetilde R, \qquad L=T - \mathcal{V},
\label{lag2}
\end{equation}
where $\widetilde R$ and $T$ are the reacting thrust and kinetic energy, given by
\begin{gather*}
\widetilde R(x,\dot x; t): = -\frac12 m'(x) \dot x^2, \quad T:= \frac12 m(x) \dot x^2.
\end{gather*}
In turn, the Hamiltonian $\mathcal{H}$ can be obtained from the Legendre transformation
\begin{gather}
\mathcal{H}(x,p;t) = p\dot x -L(x,\dot x; t) = \frac{p^2}{2m(x)} + \mathcal{V}(x).
\label{legendre}
\end{gather}
However, as the Hamiltonian's time rate of change $\dot{\mathcal{H}} = \widetilde R \dot x$ is cubic in the velocity, the variable mass system is dissipative \cite{Mus08}. That is, the Hamiltonian \eqref{legendre} is not a constant of motion. Yet, it may be shown \cite{Cru13} that the proper invariant acquires the form 
\begin{gather}
I = \frac{p^2}{2m_0} + \int^x \frac{m}{m_0} \left( \frac{\partial \mathcal{V}}{\partial r} \right) dr,
\label{inv4}
\end{gather}
where the constant $m_0$ is in mass units to get $I$ expressed in energy units. Using integration by parts in the latter result we arrive at the expression
\begin{equation}
\left( \frac{m_0}{m} \right) I = \mathcal{H}(x,p;t) - \frac{m_0}{m} \int^x \left( \frac{m}{m_0} \right)' \mathcal{V} dr.
\label{equal}
\end{equation}
Clearly, the second term at the right hand side of (\ref{equal}) is just what the Hamiltonian $\mathcal{H}$ lacks to be a constant of motion. The invariant $I$ can be expressed as a modification of the Hamiltonian (\ref{legendre}) due to an effective potential
\begin{equation}
I := \mathcal{H}_{\operatorname{ef\/f}}  \equiv \frac{m}{m_0} \left[ \frac{p^2}{2m} + \mathcal{V}_{\operatorname{ef\/f}}\right],
\label{effec}
\end{equation}
with
\begin{equation}
\mathcal{V}_{\operatorname{ef\/f}} = \mathcal{V} - \frac{m_0}{m} \int^x \left( \frac{m}{m_0} \right)' \mathcal{V} dr.
\end{equation}
For the sake of completeness let us introduce a new ``mass'' term $\mu(x) = m^2(x)/m_0$ as well as a new ``momentum'' variable $\pi = \mu \dot x$. The invariant (\ref{effec}) is simplified to $\mathcal{H}_{\operatorname{ef\/f}} = \frac{\pi^2}{2\mu} + \mathcal{V}_{\operatorname{ef\/f}}$, and the equations of motion can be expressed in conventional form
\begin{equation}
\dot x = \{ x, \mathcal{H}_{\operatorname{ef\/f}}\}_{x,\pi} = \frac{\partial \mathcal{H}_{\operatorname{ef\/f}} }{\partial \pi}, \quad \dot \pi = \{ \pi, \mathcal{H}_{\operatorname{ef\/f}}\}_{x,\pi} = -\frac{\partial \mathcal{H}_{\operatorname{ef\/f}} }{\partial x},
\label{inv6c}
\end{equation}
where the Poisson bracket
\begin{gather}
\left\{ f, g \right\}_{x, \pi} = \frac{\partial f}{\partial x} \frac{\partial g}{\partial
 \pi} -\frac{\partial f}{\partial  \pi} \frac{\partial g}{\partial x}
\label{inv7a}
\end{gather}
defines the time-variation of any smooth function $h$ depending on $x$ and $\pi$:
\begin{gather}
\frac{d}{dt} h = \left\{ h, \mathcal{H}_{\operatorname{ef\/f}} \right\}_{x, \pi} + \frac{\partial h}{\partial t}.
\label{inv7}
\end{gather}
In particular, if $h=\mathcal{H}_{\operatorname{ef\/f}}$, from (\ref{inv7}) we have $\frac{d}{dt} \mathcal{H}_{\operatorname{ef\/f}}=0$. Besides, from (\ref{inv7a}) one gets $\left\{ x, \pi \right\}_{x, \pi} = 1$, so that $x$ and $\pi$ are conjugate variables.

To summarize the results of this section let us emphasize that, although the dynamical law for a particle that suffers a spatial variation of its mass includes forces quadratic in the velocity, the Lagrangian can be written in the standard form $L=\frac{p^2}{2m(x)}-{\mathcal
V}(x)$, in correspondence with the conditions studied in~\cite{Mus08}. The construction of the related Hamiltonian also leads to the standard form ${\mathcal H}= \frac{p^2}{2m(x)}+{\mathcal V}(x)$. That is, a simple description of this kind of systems starts by replacing the (constant) mass~$m_0$ by the appropriate function of the position $m(x)$ in the conventional
expressions of~$L$ and ${\mathcal H}$. Moreover, although ${\mathcal H}$ is not time-independent, it is possible to construct an energy constant of motion $I=\mathcal{H}_{\operatorname{ef\/f}}$ leading to dynamical equations that have the form of the Hamilton ones. Further details concerning the Lagrangian and Hamiltonian formulations for PDM systems can be consulted in \cite{Cru13,Mus08}. Additional results on the matter can be found in \cite{Mus15,daC18,Hab13,Mus17,Mus19a,Mus19b}.

\subsection{Algebraic approach}

The dynamical problem~\eqref{inv6c} may be studied in two general forms \cite{Cru13}. First, given a~specif\/ic potential ${\mathcal V}(x)$ acting on the mass $m(x)$, the related phase trajectories are found. Second, given an algebra which rules the dynamical law of the mass, the potential and phase trajectories are constructed in a~purely algebraic form. The second approach has been successfully applied in previous works \cite{Cru13,Cru07,Kur08,Cru08,Cru08b,Cru13b,Cru14,Scr16,Cru18} and will be revisited in this section. The keystone is to notice that the factorization method introduced in~\cite{Cru08, Kur08} can be extended to the case of PDM classical systems that obey the dynamical law~\eqref{force}. Indeed, working in the $(x,\pi)$-plane, the factorization of the Hamiltonian~\eqref{effec} leads in a~natural form to the identif\/ication of a pair of time-dependent integrals of motion ${\mathcal Q}^{\pm}$ which, in turn, allows the construction of the phase trajectories $(x(t),\pi(t))$ associated to the canonical equations. Following \cite{Cru13}, let us look for a~couple of complex functions ${\mathcal A}^+(x,\pi;t)$, ${\mathcal A}^-(x, \pi;t)$, and a~constant $\epsilon$ such that the Hamiltonian~\eqref{effec} becomes factorized
\begin{gather}
\mathcal{H}_{\operatorname{ef\/f}} =\mathcal{A}^+\mathcal{A}^-+\epsilon=\mathcal{A}^-\mathcal{A}^++\epsilon,
\label{factor1}
\end{gather}
where
\begin{gather}
\mathcal{A}^{\pm}=\mp if(x)\frac{\pi}{\sqrt{2\mu (x)}}+g(x)\sqrt{\gamma \mathcal{H}_{\operatorname{ef\/f}}}.
\label{aes}
\end{gather}
Now, we ask the functions (\ref{aes}) to close a~deformed Poisson algebra by demanding that $\{{\mathcal A}^-,{\mathcal A}^+\}$ be expressed in terms of the powers of $\sqrt{\gamma \mathcal{H}_{\operatorname{ef\/f}}}$. In the simplest case we have 
\begin{gather}
i\left\{\mathcal{A}^-,\mathcal{A}^+\right\}=2\alpha\sqrt{\gamma \mathcal{H}_{\operatorname{ef\/f}}}, \quad
i\left\{\mathcal{H}_{\operatorname{ef\/f}},\mathcal{A}^\pm\right\}=\pm2\alpha\sqrt{\gamma \mathcal{H}_{\operatorname{ef\/f}}}\mathcal{A}^\pm,\\
\{ \mathcal{H}_{\operatorname{ef\/f}}, {\mathcal A}^+ {\mathcal A}^- \} = \{ \mathcal{H}_{\operatorname{ef\/f}}, {\mathcal A}^- {\mathcal A}^+\} = 0,
\label{palgebra}
\end{gather}
with
\begin{gather}
g(x) =
\begin{cases}
\displaystyle \sin \left[ \sqrt{2 \alpha^2m_0} \int_c^x J(t)dt \right], & \gamma=1, \ \ \  \mathcal{H}_{\operatorname{ef\/f}} >0,
\vspace{1mm}\\
\displaystyle \sinh \left[ \sqrt{2 \alpha^2m_0} \int_c^x J(t)dt \right], & \gamma=-1,  \  \mathcal{H}_{\operatorname{ef\/f}} <0,
\end{cases}
\label{g}
\end{gather}
and $J(x) =\sqrt{\mu(x)/m_0}$. The function $f$ is obtained from (\ref{g}) through $f^2(x)= 1- \gamma g^2(x)$. However, the potential allowing the above equations is not arbitrary since it depends on the $g$-function as follows
\begin{gather}
{\mathcal V}_{\text{ef\/f}} (x) = \frac{\epsilon}{1-\gamma g^2(x)} =
\begin{cases}
\displaystyle\frac{\epsilon}{\cos^2\left[ \sqrt{2 \alpha^2m_0} \int_c^x J(t)dt \right]}, & \mathcal{H}_{\operatorname{ef\/f}} >0,\\
\displaystyle\frac{\epsilon}{\cosh^2\left[ \sqrt{2 \alpha^2m_0} \int_c^x J(t)dt \right]}, & \mathcal{H}_{\operatorname{ef\/f}} <0.
\end{cases}
\label{pot}
\end{gather}
In other words, given the mass $\mu(x)$, the P\"oschl-Teller potential \eqref{pot}
is such that the factors~\eqref{aes} satisfy the deformed Poisson algebra (\ref{palgebra}). Details concerning the time-dependent integrals of motion ${\mathcal Q}^{\pm}$ as well as further properties of the potentials \eqref{pot} can be consulted in \cite{Cru13}. For other systems see, e.g., \cite{Cru08,Cru08b,Cru13b,Cru14,Scr16,Cru18}.

\section{Quantum picture}

Let us consider the one-dimensional Hamiltonian
\begin{equation}
H_a = \frac12 m^a P m^{2b} P m^a + V \equiv K_a +V, \qquad 2a+2b
=-1
\label{hamil1}
\end{equation}
where the mass $m>0$ and the potential $V$ are functions of the position, $K_a$ is the kinetic term of $H_a$ and $P$ fulfills $[X,P]=i\hbar$, with $X$ the position operator. As indicated above, the Hermiticity of the Hamiltonian $H_a$ is part of the problem to solve. In the present case the parameter $a$ defines the ordering of the mass and momentum operators, so it must be properly chosen \cite{Cru13,Cru07,Cru13b,Lev95,Mus07,Maz13}. Following \cite{Cru09,Cru13b}, the parameter $a$ is kept arbitrary, with no more assumptions on a particular ordering of $P$ and $m$. In position-representation, the eigenvalue equation
\begin{equation}
H_a \psi(x) = E\psi(x)
\label{eigen1}
\end{equation}
can be reduced to a simpler form by considering the point transformation
\begin{gather}
\psi(x) = e^{g(x)} \varphi(x), \quad x \mapsto y: =s(x),
\label{map1}
\end{gather}
where $s$ stands for a bijection that defines the Jacobian of the transformation $J:=s'(x) = \sqrt{m(x)/m_0}$, and
\begin{equation}
\int_{\operatorname{Dom} (H_a)} \vert \psi(x) \vert^2 dx = \int_{\operatorname{Dom} (H_a)} \vert \,e^{g(x)} \varphi(x) \vert^2 dx <+\infty.
\label{acota1}
\end{equation}
The straightforward calculation gives rise to the Hamiltonian
\begin{equation}
H_{\rm eff}^{(a)} \varphi (y) :=\left[-\left(
\frac{\hbar^2}{2m_0}\right) \frac{d^2}{dy^2} + V_{\rm
eff}^{(a)}(y)\right] \varphi(y) =E \varphi(y),
\label{eigen4}
\end{equation}
with an ef\/fective potential 
\begin{equation}
V_{\rm eff}^{(a)} := V- \left( \frac{\hbar^2}{2m^3} \right) \left[
\left(\frac{1}{4} +a \right) mm'' - \left\{ \frac{7}{16} +a(2+a)
\right\} (m')^2 \right]
\label{effec}
\end{equation}
that depends on the explicit expressions for the mass $m$ and the initial potential $V$, both of them in the $y$-representation. Besides, 
\begin{equation}
y =s(x) = \int e^{2g(x)} dx +y_0, \quad g(x)= \ln \left[ J^{1/2}(x) \right] = \ln \left[ \frac{m(x)}{m_0} \right]^{1/4}.
\label{gy}
\end{equation}
At this stage the main simplification is the avoiding of the mass ordering in the kinetic term of the effective Hamiltonian. Then, the techniques used to solve the eigenvalue equation of the constant mass systems may be applied to investigate the spectral problem defined by Eq.~(\ref{eigen4}). 

A further simplification is obtained if either (i) $a=-1/4$ or (ii) the mass function $m(x)$ is such that $V_{\rm eff}^{(a)} - V=0$. As the former case produces the identity $V_{\rm eff}^{(a)} = V$ for any well defined mass function $m(x)$, one says that the Hamiltonian $H_{\rm eff}^{(-1/4)}$ is defined by {\em mass-independent null terms} \cite{Cru09}. On the other hand, when the identity $V_{\rm eff}^{(a)} = V$ depends explicitly on the mass function $m(x)$, for $a\neq -1/4$ we say that $H_{\rm eff}^{(-1/4)}$ is defined by {\em mass-dependent null terms} \cite{Cru09}. In particular, a constant mass $m(x) = m_0$ reduces the effective potential (\ref{effec}) to the initial one $V$ in $y$-representation.

\subsection{Algebraic approach}

Let us factorize the Hamiltonian (\ref{hamil1}) in the form
\begin{equation}
H_a = AB + \epsilon,
\label{factor1}
\end{equation}
with $\epsilon$ a constant (in energy units) to be determined, 
\begin{equation}
A = -\frac{i}{\sqrt 2} m^a P m^b + \beta, \qquad B =
\frac{i}{\sqrt 2} m^b P m^a + \beta, \qquad A^{\dagger} = B,
\label{ab}
\end{equation}
and $\beta$ a solution of the Riccati equation
\begin{equation}
V -\epsilon = \frac{\hbar}{\sqrt{2m}} \left[ 2\left( a +
\frac{1}{4} \right) \left( \frac{m'}{m} \right) \beta - \beta'
\right] + \beta^2.
\label{riccati1}
\end{equation}
For arbitrary $m$ and $\beta$ the product between the factorization operators obeys the commutation rule:
\begin{equation}
[A,B]= -\displaystyle\left[ \frac{\hbar}{m^{3/2}} \left( a+\frac{1}{4}
\right) m'\right]^2 -\displaystyle\sqrt{\frac{2\hbar^2}{m}} \beta'.
\label{conmuta}
\end{equation}
Demanding the commutator (\ref{conmuta}) to close a given algebra we are in position of getting concrete realizations of the operators $A$, $B$. 

In the simplest case one has $[A,B]=-\hbar \omega_0$, so that the $\beta$-function is defined in terms of the ordering parameter and the mass function:
\begin{equation}
\beta = \frac{\omega_0}{\sqrt 2} \int^x m^{1/2} dr -
\frac{\hbar}{\sqrt 2} \left( a + \frac{1}{4} \right)
\left(\frac{m'}{m^{3/2}}\right) +
\beta_0.
\label{beta}
\end{equation}
Therefore
\begin{equation}
V= \frac{\omega_0^2}{2} \left[ \int^x m^{1/2} dr \right]^2 = \frac{m_0 \omega_0^2}{2} \left[ \int^x J(r) dr \right]^2
\label{pot2}
\end{equation}
Notice that $m(x)=m_0$ produces the harmonic oscillator potential $V(x)= \frac{m_0 \omega_0^2}{2}  x^2$, with $\beta(x) = \left( \frac{m_0 \omega_0^2}{2} \right)^{1/2} x + \beta_0$. Then, up to an additive constant, the operators $A$ and $B$ are reduced to the conventional ladder operators of the harmonic oscillator, as expected. For other forms of the mass function $m(x)$ and the appropriate ordering parameter $a$, the potential (\ref{pot2}) represents a wide family of PDM potentials with the energy spectrum of the harmonic oscillator  \cite{Cru09}. On the other hand, the introduction of (\ref{beta}) into (\ref{ab}) generates the ladder operators for such PDM oscillators. The construction of the corresponding generalized coherent states is also feasible \cite{Cru09}.

Other PDM quantum systems can be studied through the commutator (\ref{conmuta}) by identifying the appropriate algebra. For instance, we may look for operators $A$ and $B$ such that the commutator (\ref{conmuta}) is associated with the $su(1,1)$ Lie algebra. The potential (\ref{pot2}) is in such case associated with a family of singular oscillators \cite{Cru11}. Additional constructions of coherent states for PDM systems have been reported in Refs.~\cite{Bis09,Rub10,Yah12,Yah14,Ami16a,Ami16b,Ami17,Yah17,Ros19}. Further details concerning the properties of quantum systems with position-dependent mass can be consulted in, e.g., \cite{Guo09,Mus10,Mus11,Lim12,Bag13a,Mus13,Bag13b,Cho13,Vub14,daC14,Bag15,Gho15,Mol16,Che17,Nik17,Nik19,Mus19c,Oli19,Mus20}.

\subsection*{Acknowledgment}
The author is indebted to Sara Cruz~y~Cruz, with whom most of the results reported in this work have been obtained. The support from Consejo Nacional de Ciencia y Tecnolog\'ia (Mexico), grant number A1-S-24569, is acknowledged.



\end{document}